\documentclass[preprint]{aastex}

\usepackage{lineno}
\usepackage{graphicx}

\newcommand{\msol}{\hbox{$M_\odot$}}                   

\makeatletter

\newcommand{\Rmnum}[1]{\expandafter\@slowromancap\romannumeral #1@}
\makeatother

\begin{document}

\title{Fermi-LAT and WMAP observations of the Puppis A Supernova Remnant}

\author{
J.W.~Hewitt\altaffilmark{2,1}, 
M.-H.~Grondin\altaffilmark{3,4,1}, 
M.~Lemoine-Goumard\altaffilmark{5,6,1}, 
T.~Reposeur\altaffilmark{5}, 
J.~Ballet\altaffilmark{7} and 
T.~Tanaka\altaffilmark{8}
}
\altaffiltext{1}{Corresponding authors: M.-H.~Grondin, marie-helene.grondin@mpi-hd.mpg.de; J.W.~Hewitt, john.w.hewitt@nasa.gov; M.~Lemoine-Goumard, lemoine@cenbg.in2p3.fr.}
\altaffiltext{2}{NASA Goddard Space Flight Center, Greenbelt, MD 20771, USA}
\altaffiltext{3}{Max-Planck-Institut f\"ur Kernphysik, D-69029 Heidelberg, Germany}
\altaffiltext{4}{Landessternwarte, Universit\"at Heidelberg, K\"onigstuhl, D 69117 Heidelberg, Germany}
\altaffiltext{5}{Universit\'e Bordeaux 1, CNRS/IN2p3, Centre d'\'Etudes Nucl\'eaires de Bordeaux Gradignan, 33175 Gradignan, France}
\altaffiltext{6}{Funded by contract ERC-StG-259391 from the European Community}
\altaffiltext{7}{Laboratoire AIM, CEA-IRFU/CNRS/Universit\'e Paris Diderot, Service d'Astrophysique, CEA Saclay, 91191 Gif sur Yvette, France}
\altaffiltext{8}{W. W. Hansen Experimental Physics Laboratory, Kavli Institute for Particle Astrophysics and Cosmology, Department of Physics and SLAC National Accelerator Laboratory, Stanford University, Stanford, CA 94305, USA}

\keywords{cosmic rays -- acceleration of particles -- ISM: individual (Puppis A) -- radiation mechanisms: nonthermal }

\slugcomment{Draft prepared on \today }

\begin{abstract}
We report the detection of GeV $\gamma$-ray emission from the supernova remnant Puppis A with the {\it Fermi Gamma-Ray Space Telescope}. Puppis A is among the faintest supernova remnants yet detected at GeV energies, with a luminosity of only $2.7 \times$10$^{34}$ ($D / 2.2$ kpc)$^2$ erg s$^{-1}$ between 1 and 100 GeV. The $\gamma$-ray emission from the remnant is spatially extended, with a morphology matching that of the radio and X-ray emission, and is well described by a simple power law with an index of 2.1. We attempt to model the broadband spectral energy distribution, from radio to $\gamma$-rays, using standard nonthermal emission mechanisms. To constrain the relativistic electron population we use 7 years of WMAP data to extend the radio spectrum up to 93 GHz. Both leptonic and hadronic dominated models can reproduce the nonthermal spectral energy distribution, requiring a total content of cosmic ray (CR) electrons and protons accelerated in Puppis~A of at least $\rm W_{CR} \approx$ (1 -- 5)$\times 10^{49}$ erg.
\end{abstract}

\maketitle

\section{Introduction}\label{sec:introduction}

Supernovae have long been thought responsible for accelerating protons to relativistic energies in our Galaxy. Diffusive shock acceleration provides a mechanism to convert a significant fraction of the blast wave kinetic energy into relativistic particles. However it has proven difficult to measure cosmic rays (CRs) from supernova remnants (SNRs) directly \citep[and references therein]{Reynolds2008}.

The {\it Fermi Gamma-Ray Space Telescope} ({\it Fermi}) has identified a number of SNRs as high-energy $\gamma$-ray sources. Middle-aged SNRs interacting with molecular clouds are among the most luminous GeV sources in the Galaxy \citep{fermi:w51c,fermi:w28,fermi:w44,fermi:ic443,castro2010}. The high luminosities are thought to result from interactions between cosmic rays and the large target mass of the molecular clouds, though it is not yet certain whether GeV emission arises from leptonic (Bremsstrahlung) or hadronic ($\pi^0$ decay) processes. In contrast, the young SNR Cassiopeia A (hereafter Cas A), is detected with a luminosity two orders of magnitude lower than SNRs interacting with molecular clouds \citep{fermi:casa}. A careful consideration of $\gamma$-ray emission mechanisms in Cas A indicates that at least 6$\times$10$^{49}$ erg in hadronic CRs are required \citep{araya10}. Other young SNRs have also been detected by $Fermi$ -- Tycho \citep{fermi:tycho}, RX J1713.7$-$3946 \citep{fermi:rxj1713}, and RX J0852.0--4622 \citep{fermi:velajr} -- though the origin of $\gamma$-ray emission remains uncertain. Clearly, SNRs contribute at least part of the Galactic CR population, though the total energetic input and timescale for acceleration require further study.

Supernova remnant Puppis A (G260.4--3.4) is an important case to study, as it shows signs of recently encountering a higher ambient density in the vicinity of a nearby molecular cloud \citep{Hwang2005}. The remnant displays increasing X-ray surface brightness from west to east \citep{petre82} corresponding to an increasing density of the ambient interstellar medium (ISM) at the eastern and northern shell \citep{Dubner1988}. The proper motions of fast optical knots gives a dynamical age of 3700$\pm$300 years, establishing that the SNR is in the Sedov-Taylor evolutionary phase \citep{Winkler1988}. X-ray spectra from the rim of Puppis A show low elemental abundances relative to solar, consistent with swept-up interstellar medium (ISM) and a lack of ejecta contamination \citep{Katsuda2008}. A progenitor mass of 15--25 \msol\ is inferred from comparisons of relative abundances of metal-rich ejecta with nucleosynthesis models \citep{Hwang2008, Katsuda2010}. A plausible SN \Rmnum{2}L/b progenitor would have produced a clumpy, red supergiant wind only out to 7 pc \citep{Chevalier2005}. At an estimated distance of 2 kpc \citep{Reynoso1995}, the diameter of Puppis~A is 30 pc, so the remnant is no longer interacting with the circumstellar medium of the progenitor, but the surrounding ISM.

The extent of the shock interaction is characterized by thermal dust emission in infrared (IR). \cite{Arendt1991} noted enhanced dust emission corresponding with the bright X-ray shell. The dust emission is approximated by a 65 K black body, whereas thermal X-rays originate from a $>$10$^6$ K gas \citep{Hwang2005}. {\it Spitzer Space Telescope} imaging of the SNR at 24, 70 and 160 $\mu$m reveals a detailed correlation between IR and X-ray emission on arcsecond scales \citep{Arendt2010}. This is interpreted as swept-up interstellar dust heated by collisions with the hot shocked plasma observed in X-rays. There are two notable regions where the SNR shock has engulfed small denser clouds: the Bright Eastern Knot and the Northern Knot. Besides those the shock has not yet become radiative for most of the SNR, consistent with the relatively young age and the low density of the surrounding medium. 

Here we report the detection of high energy $\gamma$-ray emission from 200 MeV to 100 GeV, based on observations with the Large Area Telescope (LAT), the primary science instrument on $Fermi$. We also analyze the WMAP 7-year data from 23 to 93 GHz, in order to better constrain the relativistic electron population. The excellent characterization of the physical properties of the Puppis A makes it an ideal candidate to explore the production of nonthermal emission this SNR.

\section{Observations and Data Analysis}\label{sec:results}\label{sec:observations}

\subsection{$Fermi$-LAT observations and data reduction}
\label{ana}
The LAT detects $\gamma$-ray photons by conversion into electron-positron pairs in the energy range between 20 MeV to higher than 300 GeV, as described by~\cite{atwood09}.
It contains a high-resolution converter/tracker (for direction measurement of the incident $\gamma$-rays), a CsI(Tl) crystal calorimeter (for energy measurement), and an anti-coincidence detector to identify the background of charged particles. 
The LAT has a large effective area ($\sim$ 8000 cm$^{2}$ on-axis above 1~GeV), 
a wide field of view ($\sim$ 2.4 sr) and good angular resolution ($\sim$0.6$^{\circ}$ radius for 68$\%$ containment at 1 GeV for events converting in the front section of the tracker). The on-orbit calibration is described in \cite{OnorbitCalib}. 

The following analysis was performed using 36 months of data collected from 2008 August 4 to 2011 August 20 within a $15\degr \times 15\degr$ region around the position of Puppis A. Only events with Earth zenith angles smaller than 100$\degr$ were included to reduce contamination from the Earth limb. We used the P7V6 instrument response functions (IRFs), and selected the `Source' events which correspond to the best compromise between the number of selected photons and the charged particle residual background for the study of point-like or slightly extended sources. 

Two different tools were used to perform the spatial and spectral analysis: $\mathtt{gtlike}$ and $\mathtt{pointlike}$. $\mathtt{gtlike}$ is a binned maximum-likelihood method \citep{mattox96} implemented in the Science Tools distributed by the $Fermi$ Science Support Center (FSSC)\footnote{More information about the performance of the LAT can be found at the FSSC ($\mathtt{http://fermi.gsfc.nasa.gov/ssc}$).}.  $\mathtt{pointlike}$ is an alternate binned likelihood technique, optimized for characterizing the extension of a source (unlike $\mathtt{gtlike}$), that was extensively tested against $\mathtt{gtlike}$ \citep{kerrphd, fermi:extsrccat}. These tools fit a source model to the data along with models for the residual charged particles and diffuse $\gamma$-ray emission. In the following analysis, the Galactic diffuse emission is modeled by the standard LAT diffuse emission ring$-$hybrid model \emph{gal\_2yearp7v6\_v0.fits}. The residual background and extragalactic radiation are described by a single isotropic component with a spectral shape described by the file \emph{iso\_p7v6source.txt}. These models are available from the FSSC.

Sources within $15\degr$ around Puppis A listed in the $Fermi$-LAT Second Source Catalog \citep[hereafter 2FGL]{fermi:2fgl} are included in our spectral-spatial model of the region. We also include extended spatial templates describing the $\gamma$-ray emission from RX J0852.0--4622 \citep{fermi:velajr} and the pulsar wind nebula (PWN) Vela-X \citep{fermi:velaX2}. Puppis A is associated with three 2FGL sources within the radio boundary: 2FGL J0821.0--4254, J0823.0--4246, J0823.4--4305. These sources are either all left free to vary or are replaced by other spatial models. The spectral parameters of sources closer than 5$^{\circ}$ to Puppis A are left free, while the parameters of all other sources are fixed at the 2FGL values. 

Located at  a distance of  3$^{\circ}$ from the SNR Puppis~A, the Vela pulsar is the brightest steady $\gamma$-ray source in the sky, from which photons are observed up to 25 GeV~\citep{fermi:vela2}. To avoid any bias on the analysis of the SNR due to this bright nearby pulsar, all studies involving events below 3~GeV were performed in the off-pulse window of the Vela pulsar.
Gamma-ray photons were phase-folded using a timing solution produced from observations made with the Parkes 64 m radio telescope~\citep{Weltevrede2010}. To build this timing solution, 163 times of arrival (TOA)  were used covering the period from the beginning of the mission to 2011 August 20. We fit the radio TOAs to the pulsar rotation frequency and first ten derivatives. The fit further includes 15 harmonically related sinusoids, using the ÒFITWAVESÓ option in the TEMPO2 package \citep{tempo2}, to flatten the timing noise. The post-fit rms is 119.579 $\mu$s, or 0.13\% of the pulsar phase.  This timing solution will be made available through the FSSC\footnote{Public page for pulsar timing models: http://fermi.gsfc.nasa.gov/ssc/data/access/lat/ephems/}. 

The off-pulse phase window was chosen between 0.65 -- 1.05, slightly larger than the one used for the analysis of the Vela-X PWN~\citep{fermi:velaX} allowed by the larger angular distance between Puppis~A and Vela than between the pulsar and its associated PWN. This off-pulse choice offers good statistics without increasing the systematics due to the contamination from the Vela pulsar.

\subsection{Morphological analysis of $Fermi$-LAT data}
\label{like}
In the study of the morphology of an extended source, one of the major objectives is to have the best possible angular resolution. Therefore, we restrict our LAT dataset to events with energies above 800 MeV. This also reduces the relative level of the Galactic diffuse background owing to its softer spectrum. As discussed above, to ensure that we do not suffer contamination from the Vela pulsar, we only selected the off-pulse photons defined as having phases 0.65 -- 1.05.

\begin{figure}
\centering
\includegraphics[height=4in]{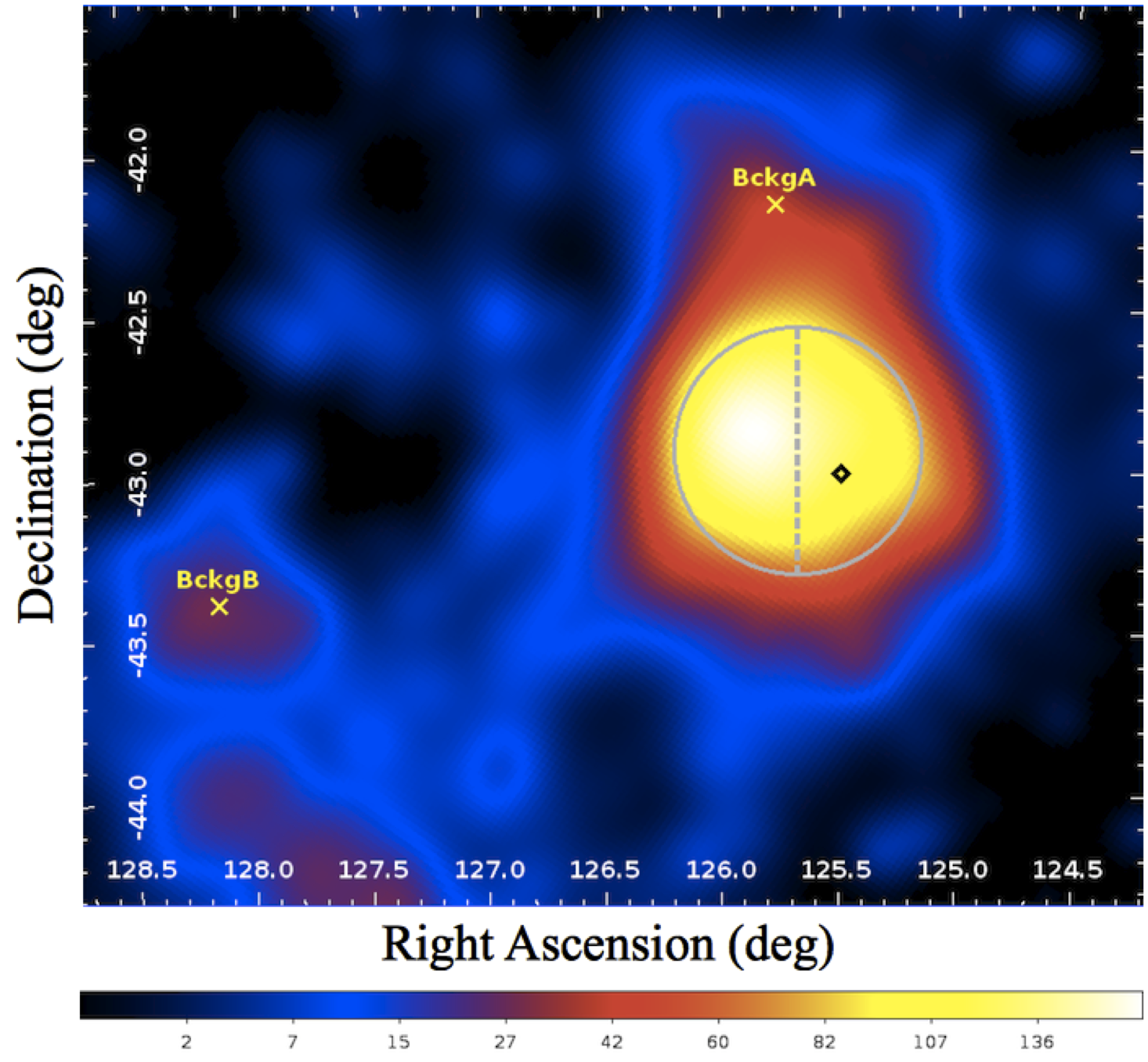}
\caption{Test Statistic (TS) map in celestial coordinates of the SNR Puppis~A. Only photons with energies above 800 MeV in the off-pulse window of the Vela pulsar are selected. The TS was evaluated by placing a point-source at the center of each pixel, Galactic diffuse emission and nearby sources being included in the background model, except the two background sources BckgA and BckgB which are indicated with yellow crosses. The location of the compact object inside Puppis~A, PSR J0821$-$4300, is indicated by a black diamond. The gray circle corresponds to the best fit model ($D1$) provided by $\mathtt{pointlike}$ in Table~\ref{tab:pointlike}, the dashed line showing its division in two half-disks along the North/South line in celestial coordinates. The color-coding is represented on a square-root scale.
\label{fig:tsmap} }
\end{figure}

We fit models to the data using the maximum likelihood framework. The test statistic (TS) of a source is defined as twice the logarithm of the ratio between the likelihood $\mathcal{L}_1$ obtained by fitting the source model plus background components (including other sources) to the data, and the likelihood $\mathcal{L}_0$ obtained by fitting the background components only, i.e TS = 2 log($\mathcal{L}_1$/$\mathcal{L}_0$). 
Figure~\ref{fig:tsmap} shows a map of the point-source detection significance, evaluated at each point in the TS map for the region around Puppis A using photons with energies $>$ 800 MeV. This skymap contains the TS value for a point source of fixed photon index $\Gamma = 2$ at each map location, thus giving a measure of the statistical significance for the detection of a $\gamma$-ray source with that spectrum in excess of the background. Emission coincident with Puppis A is detected, but also in two regions outside of its shell. We consider these regions of excess $\gamma$-ray emission as likely to belong to background sources not recognized in 2FGL. Due to the longer integration time of our analysis (36 months vs. 24 months in the catalog) and the overwhelming brightness of the Vela pulsar in the entire phase interval, the appearance of additional sources in our region of interest is expected. These sources are denoted with the identifiers BckgA and BckgB and will be described below.

\begin{figure}
\centering
\includegraphics[height=2.1in]{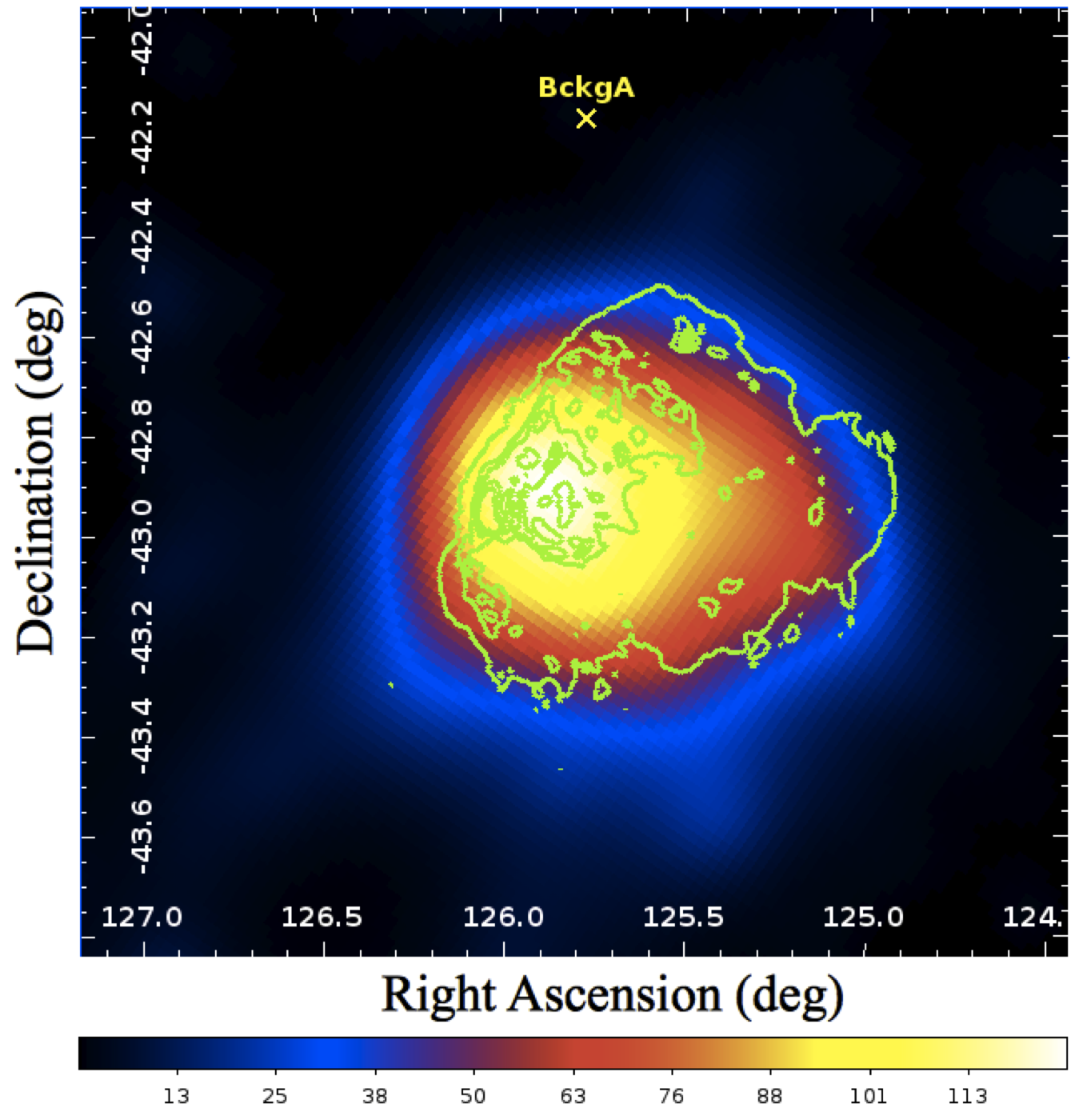}
\includegraphics[height=2.1in]{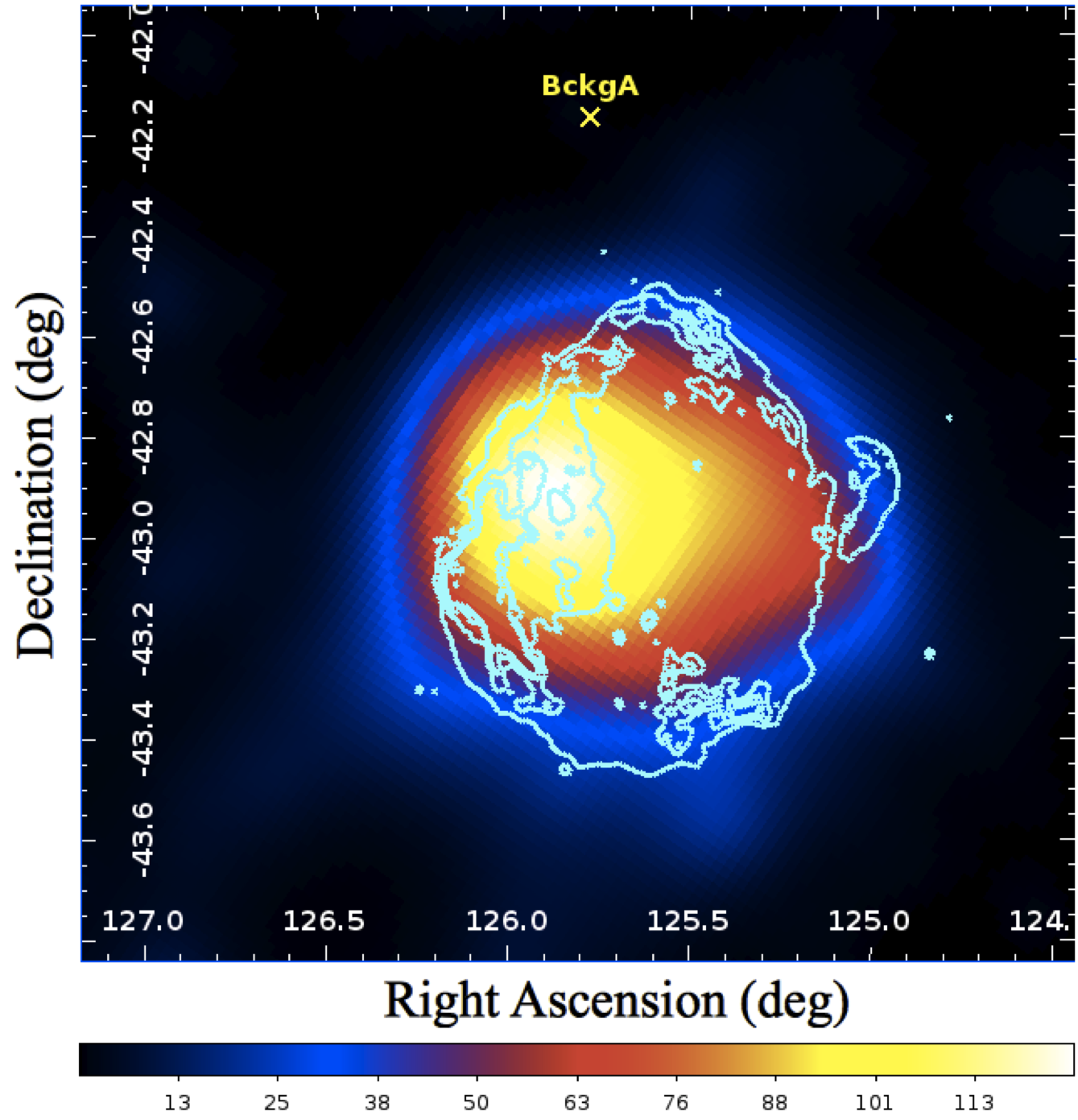}
\includegraphics[height=2.1in]{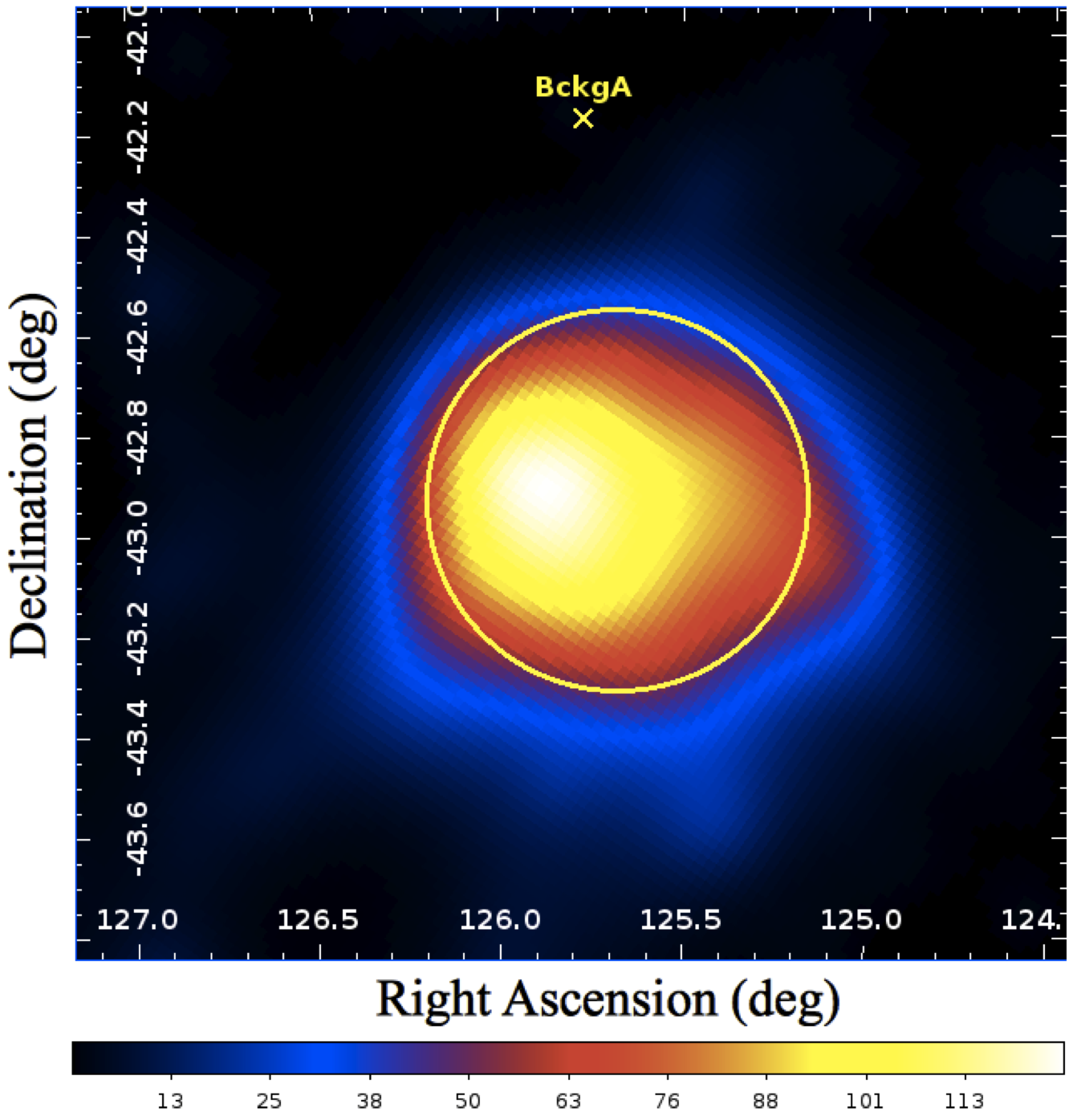}
\vspace{0.3cm}
\includegraphics[height=2.1in]{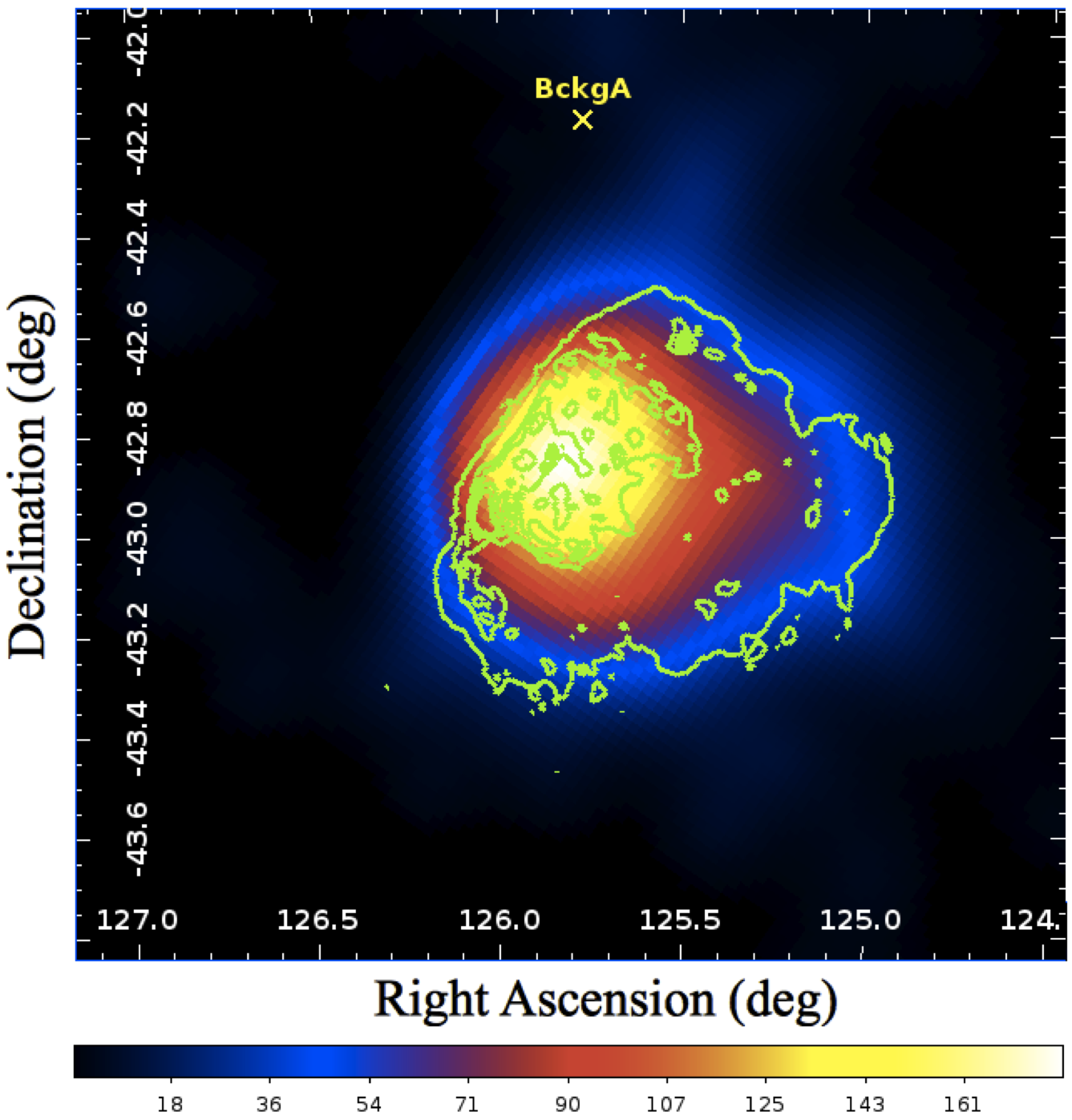}
\includegraphics[height=2.1in]{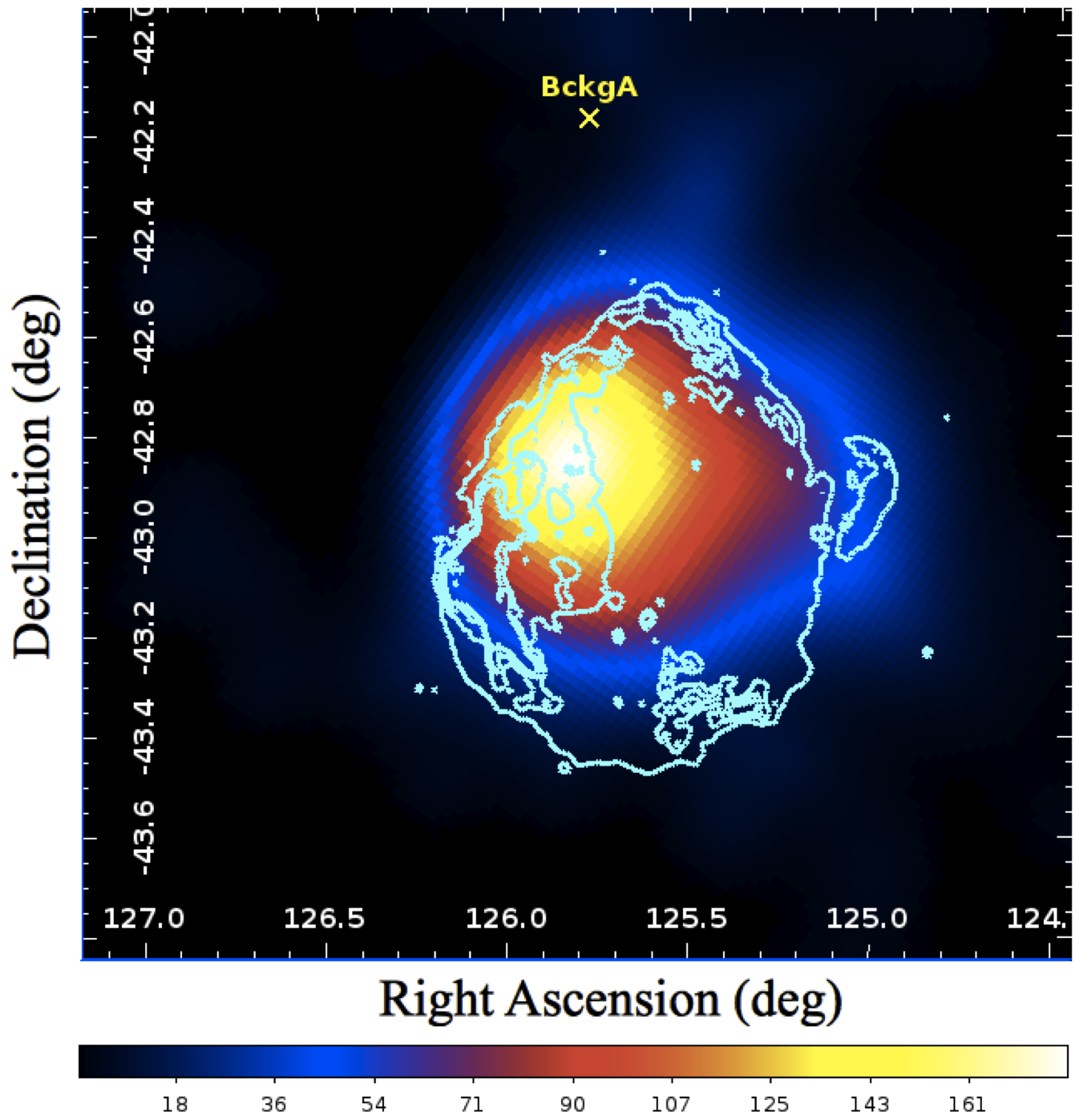}
\includegraphics[height=2.1in]{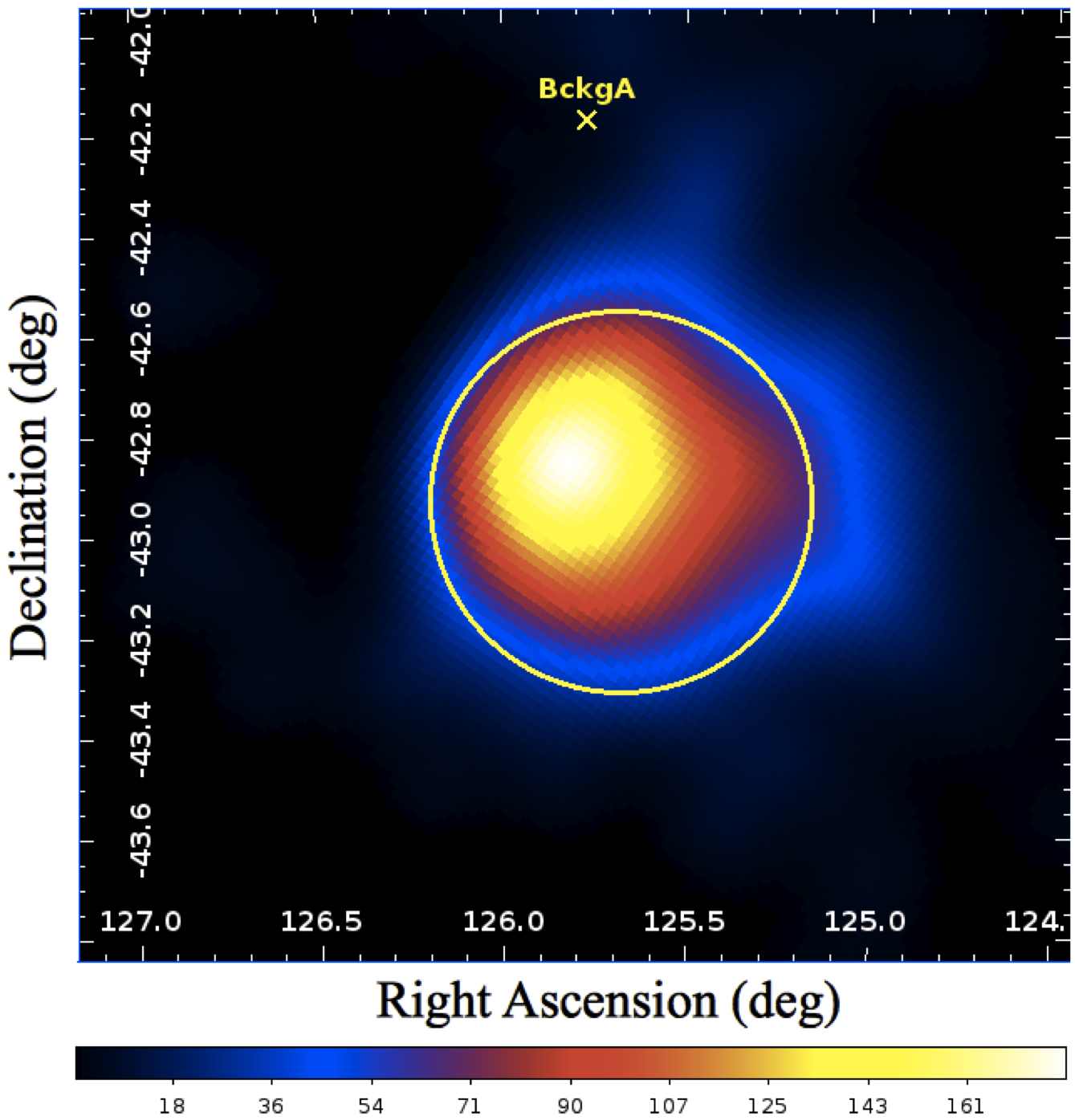}
\caption{Test Statistic (TS) map of the SNR Puppis~A above 800 MeV in the off-pulse window of the Vela pulsar (Top) and above 5 GeV in the all-phase interval (Bottom). The TS was evaluated by placing a point-source at the center of each pixel, Galactic diffuse emission and nearby sources being included in the background model. The location of the background source BckgA added to the source model is indicated with a yellow cross. Green and cyan contours correspond to images at different wavelengths: ROSAT HRI X-ray emission contours (Green, Left) and 1.4 GHz radio continuum contours (Cyan, Middle). The contour levels are 0.02, 0.19, 0.37, 0.55 counts s$^{-1}$ m$^{2}$ for X-ray emission, and 5, 10, 15, 20 mJy beam$^{-1}$ for 1.4 GHz radio continuum. The yellow disk corresponds to the best fit model ($D1$) provided by $\mathtt{pointlike}$ in Table~\ref{tab:pointlike}. The color-coding is represented on a linear scale for all images.}
\label{fig:tsmaps_zoom} 
\end{figure}

Puppis A was first established as an extended source above 1 GeV by \cite{fermi:extsrccat}. 
Following their procedure, we determined the source extension using $\mathtt{pointlike}$ with a uniform disk hypothesis (compared to the point-source hypothesis) for energies above 800 MeV in the off-pulse of the Vela pulsar and above 5 GeV in the all-phase interval. The results are summarized in Table~\ref{tab:pointlike}. To quantify the significance of the extension of the SNR Puppis~A, we define TS$_{ext}$ as twice the difference between the log-likelihood of an extended source model and the log-likelihood of a point-like source model. Above 800 MeV, TS$_{ext}$=46 (which converts into a significance of $\sim 7 \sigma$ for the source extension), demonstrating that the source is significantly extended with respect to the LAT point-spread function (PSF). The fitted radius is $0.38^{\circ} \pm 0.04^{\circ}$, in good agreement with the size of the SNR as seen in the radio and X-rays. The extension is still significant above 5 GeV with a slightly smaller size of $0.32^{\circ} \pm 0.03^{\circ}$. Going from lower to higher energy, the radius of the SNR seems to decrease with a fitted position centered on the Eastern side of the remnant but, due to the large uncertainty of the fit with the current statistics, this effect is only marginal. This trend is visible in Figure~\ref{fig:tsmaps_zoom} which presents the LAT TS maps above 800 MeV and above 5 GeV when nearby sources and the extragalactic and Galactic components of the background are included in the background model. The emission above 5 GeV is brightest in the East, near the bright eastern knot, where the SNR shock is interacting with a dense cloud, and similar to the radio and X-ray morphologies.

It should be noted that during the fit above 800 MeV, we simultaneously localized the sources BckgA and BckgB assuming the best disk spatial model for the SNR Puppis~A, presented in Table~\ref{tab:pointlike}. We followed the procedure described in \cite{fermi:extsrccat}. BckgA was fit to position $\alpha(\rm{J2000})=125.77\degr$ , $\delta(\rm{J2000})=-42.17\degr$ with a 68\% error radius of 0.06$\degr$, while BckgB was fit at $\alpha(\rm{J2000})=128.14\degr$, $\delta(\rm{J2000})=-43.39\degr$ with a 68\% error radius of 0.05$\degr$. The addition of sources BckgA and BckgB in the source model results in an improved maximum likelihood above 800 MeV with four additional degrees of freedom. The sources have corresponding significances of $\sim$~4.4 and $\sim$~5.5$\sigma$ (TS = 27 and 39) respectively, demonstrating that these two sources are significant and distinct from Puppis A. In addition, we will show in \S~\ref{spec} that, in a combined likelihood analysis of the spectrum of Puppis~A, the emission from these two sources is much softer than the $\gamma$-ray emission from the SNR. 

To be more quantitative, we have examined the correspondence of the $\gamma$-ray emission from Puppis~A with different source morphologies by using $\mathtt{gtlike}$ with assumed multi-frequency templates above 800 MeV. For this exercise we compared the TS of the best-fit uniform disk model provided by $\mathtt{pointlike}$ (see Table~\ref{tab:pointlike}) with TS derived when using morphological templates from ROSAT and the VLA (1.4 GHz radio images) assuming a power-law spectrum. The resulting TS values obtained from our maximum likelihood fitting are summarized in Table~\ref{tbl:spatial_models}. The radio template, the X-ray template and the uniform disk all produce improvements from the 2FGL 3-point source model, while having fewer degrees of freedom. Finally, we divided the ROSAT and uniform disk templates into two half-disks along the North/South line in celestial coordinates, as indicated in Figure~\ref{fig:tsmap}, to verify the evidence from $\mathtt{pointlike}$ of an energy-dependent shape. The two-half-disk model provides a marginal improvement ($\sim2.4 \sigma$ level), in the same trend described above. However, further data are needed to confirm this effect.

\begin{deluxetable}{lcccccc}
\tablecaption{Centroid and extension fits to the LAT data for Puppis~A using $\mathtt{pointlike}$ for events with energies between 800~MeV and 100 GeV and between 800 MeV and 5 GeV in the off-pulse window of the Vela pulsar, and above 5 GeV in the all-phase interval. \label{tab:pointlike} }
\tablewidth{0pt}
\tablehead{ 
\colhead{Spatial Model} & \colhead{Name} & \colhead{Energy (GeV)} & \colhead{R.A. ($^\circ$)} & \colhead{Dec. ($^\circ$) } & Radius ($^\circ$) & \colhead{TS$_{ext}$} 
}
\startdata
Point Source 	& $PS1$ 	& 0.8 -- 100	& 125.85   & $-$42.90 	&  & \\
Disk			& $D1$ 	& 0.8 -- 100	& 125.67 	& $-$42.93 	& 0.38 $\pm$ 0.04 & 46 \\
Point Source 	& $PS2$ 	& 0.8 -- 5	& 125.62 	& $-$42.92	&  & \\
Disk			& $D2$ 	& 0.8 -- 5	& 125.71	&  $-$42.85	& 0.47 $\pm$ 0.08 & 14 \\
Point Source 	& $PS3$ 	& 5 -- 100	&  125.83	& $-$42.87 	&  & \\
Disk			& $D3$ 	& 5 -- 100	&  125.69 & $-$42.90 	& 0.32 $\pm$ 0.03 &  26 \\
\enddata
\end{deluxetable}

\begin{deluxetable}{lcc}
\tablecaption{Test statistic obtained using $\mathtt{gtlike}$ for different spatial models compared with the null hypothesis of no
$\gamma$-ray emission associated with the SNR Puppis~A. The photon energies are 0.8 -- 100 GeV; only off-pulse photons are selected. For each model, we give the name, the difference in log-likelihood value and the number of additional degrees of freedom (Ndf) \label{tbl:spatial_models}}
\tablewidth{0pt}
\tablehead{ 
   \colhead{Spatial Model} & \colhead{TS} & \colhead{Ndf} }
\startdata
Null hypothesis & 0 & 0 \\
Point Source ($PS1$)			& 120 & 4	\\
Three Point Sources (2FGL)	& 160 & 12 	\\
Radio Template				& 166 & 2 	\\
X-ray Template				& 170 & 2	\\
Uniform Disk ($D1$)			& 172 & 5	\\
Two Half-Disk (E, W)	& 180 & 7	\\
Two X-ray halves (E, W)	& 178 & 4	\\
\enddata
\end{deluxetable}

\subsection{Spectral analysis of  $Fermi$-LAT data}
\label{spec}
Using the different templates described above, we performed maximum likelihood fits and compared the best-fit parameters in the wide energy range 0.2--100 GeV. To avoid any bias due to the brightness of the Vela pulsar, the spectral fit was performed in the off-pulse phase interval. No evidence for cutoff or break is visible and, in all cases, the $Fermi$-LAT data are well described by
a power-law such that the differential photon flux, $F(E)$, is given by:

\begin{equation}
\label{EQ:dNdE}
  F(E)\,=\,\frac{dN}{dE}\,=\,F_0\,\left(\frac{E}{E_0}\right)^{-\Gamma}
\end{equation}

\noindent where $F_0$ is the differential flux at energy $E_0$ and $\Gamma$ is the photon index. The uncertainty contours as a function of energy $E$, called the butterfly, are defined such that the differential flux satisfies:

\begin{equation}
\label{EQ:dFF}
\frac{{\Delta}F^2}{F^2}\,=\,\left(\frac{{\Delta}F_0}{F_0}\right)^2\,+\,\textrm{ln}^2\left(\frac{E}{E_0}\right)\Delta\Gamma^2\,-\,\frac{2}{F_0}\,\textrm{cov}(F_0,\Gamma)\,\textrm{ln}\left(\frac{E}{E_0}\right)
\end{equation}

\noindent where cov($F_0,\Gamma$) is the covariance term, returned by
the $MINUIT$ minimization and error analysis function called by the $Fermi$-LAT likelihood analysis tool, 
$\mathtt{gtlike}$, and ${\Delta}F$, ${\Delta}F_0$ and ${\Delta}\Gamma$ are the statistical uncertainties
on $F$, $F_0$ and $\Gamma$, respectively.

As shown in Table~\ref{tbl:spectra}, the obtained fluxes and spectral indexes are almost the same for all spatial templates used. 
These spectral fits take into account the $\gamma$-ray emission of the two additional background sources BckgA and BckgB. Their positions and spectral parameters (fitted using the uniform disk $D1$ to describe Puppis~A) are given in Table~\ref{tbl:spectra2}.

\begin{deluxetable}{lccc}
\tablecaption{Best-fit spectral parameters obtained with $\mathtt{gtlike}$ using different templates for Puppis~A above 200 MeV. The source model includes the nearby sources BckgA and BckgB described in Table~\ref{tbl:spectra2}. Only off-pulse photons from the Vela pulsar are selected and fluxes are not renormalized to the whole phase interval. The first and second errors denote statistical and systematic errors, respectively (see \S~\ref{spec})\label{tbl:spectra}}
\tablewidth{0pt}
\tablehead{
   \colhead{Spatial Model} & \colhead{Flux($>$200 MeV)} & \colhead{Photon index} & \colhead{TS}\\ & \colhead{[$10^{-8}$ ph cm$^{-2}$ s$^{-1}$]}}
\startdata
X-ray & $1.64 \pm 0.23 \pm 0.21$ & $2.09 \pm 0.07 \pm 0.09$ & 189 \\
Radio & $1.64 \pm 0.22 \pm 0.22$ & $ 2.06 \pm 0.07 \pm 0.08$ & 187 \\
Uniform disk ($D1$) & $1.67 \pm 0.23 \pm 0.23$ & $2.10 \pm 0.07 \pm 0.10$ & 190 \\
Two X-ray halves:      &  & \\
Eastern half-disk   & $0.68 \pm 0.24 \pm 0.17$ &  $1.96 \pm 0.13 \pm 0.08$ & 82 \\
Western half-disk & $0.94 \pm 0.29 \pm 0.16$ & $2.28 \pm 0.14 \pm 0.10$ & 54 \\ 
\enddata
\end{deluxetable}

\begin{deluxetable}{lccccc}
\tablecaption{Best-fit spectral parameters obtained on the nearby sources BckgA and BckgB with $\mathtt{gtlike}$ using the uniform disk template $D1$ for Puppis~A above 200 MeV. Only off-pulse photons from the Vela pulsar are selected and fluxes are not renormalized to the whole phase interval. The first and second errors denote statistical and systematic errors, respectively (see \S~\ref{spec})\label{tbl:spectra2}}
\tablewidth{0pt}
\tablehead{
   \colhead{Source Name} & \colhead{R.A. ($^{\circ}$)} & \colhead{Dec. ($^{\circ}$)} & \colhead{Flux($>$200 MeV)} & \colhead{Photon index} & \colhead{TS}\\ & & & \colhead{[$10^{-8}$ ph cm$^{-2}$ s$^{-1}$]}}
\startdata
BckgA & 125.77 & $-$42.17 & $1.17 \pm 0.26 \pm 0.23$ & $2.62 \pm 0.13 \pm 0.15$ & 48\\
BckgB & 128.14 & $-$43.39 & $1.72 \pm 0.24 \pm 0.20$ & $2.60 \pm 0.11 \pm 0.12$ & 89\\
\enddata
\end{deluxetable}

We searched for spectral variations across the $\gamma$-ray emission associated with Puppis~A. We allowed an independent normalization and spectral index for the two X-ray halves. Interestingly, the spectrum of the Western half-disk is found to be steeper by an index of 0.3 ($2 \sigma$) above 200 MeV, with respect to the Eastern half-disk. However, more statistics are needed to confirm spectral differences between the Eastern and Western regions. The spectral parameters for each of the two half-disks of the remnant are reported for reference in Table~\ref{tbl:spectra}. Note that the TS of each individual half is computed by removing only that half, so the total likelihood improvement from the two X-ray halves model (Table 2) is more than the sum of the TS of the individual halves (Table 3).

To ensure that the slightly softer spectrum in the Western half-disk is not
due to contamination from the compact object inside the SNR Puppis~A, PSR~J0821$-$4300, we performed a timing analysis using the same $Fermi$-LAT dataset. The low TS of this side of the remnant combined with the need to keep
only photons in the Vela pulsar off-pulse make a blind pulsation search unfeasible. We used ephemerides from \cite{Gotthelf2009} with the assumption of a zero period derivative \citep{Gotthelf2010} to fold the
photon arrival times using the $Fermi$ plugin distributed with the TEMPO2 software. We applied the {\it H}-test periodicity test \citep{deJager2010} for several cuts in both energy and photon distance from the pulsar location and found no evidence of pulsation with a significance better than {\it H}-test=13.5 (2.8$\sigma$). This confirms that with a small spin-down luminosity and a relatively weak
magnetic field \citep{Gotthelf2010} PSR~J0821$-$4300 is not a good candidate for $\gamma$-ray emission.

\begin{figure}
\centering
\includegraphics[width=6in]{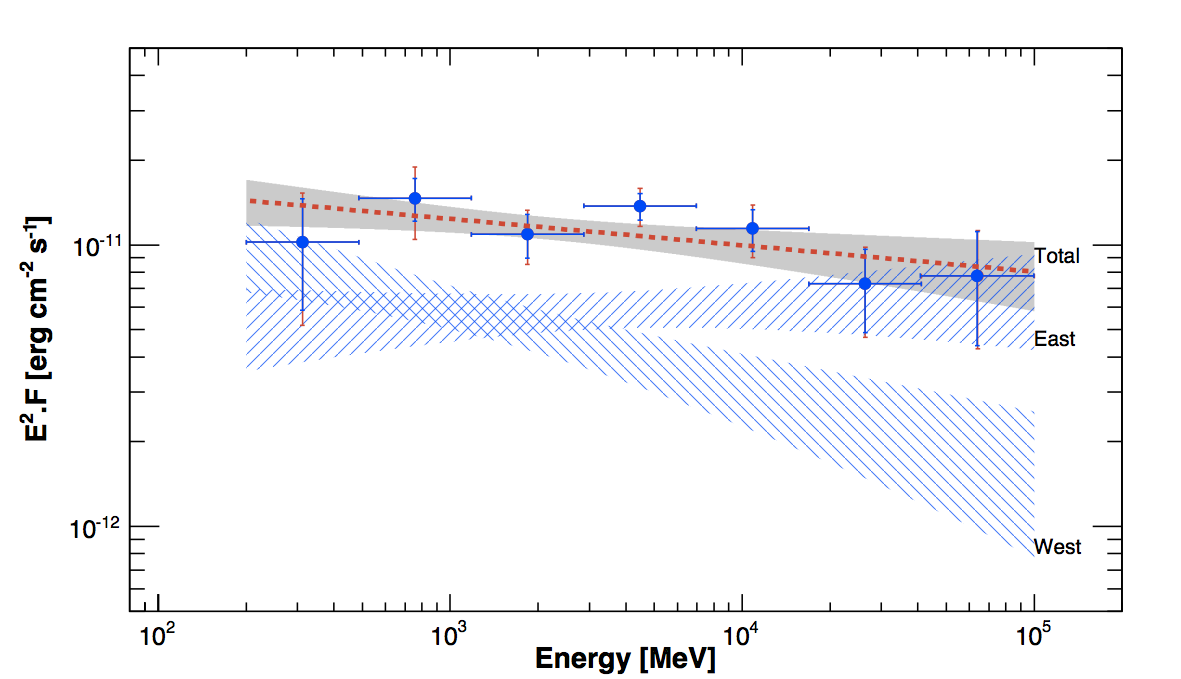}
\caption{$Fermi$-LAT spectrum of Puppis A using the X-ray spatial model described in the text. The bold-dashed (red) line shows the fit of a power law to the overall spectrum derived from all of the data with energy E $>$ 200 MeV. The data points (crosses) indicate the flux measured in each of the seven energy bins renormalized to the whole phase interval, as indicated by the extent of their horizontal lines. The statistical { 1$\sigma$} errors are shown in dark (blue), while the light (red) lines take into account both the statistical and systematic errors as discussed in \S~\ref{spec}. The gray shaded area shows the extent of the Fermi 68\% confidence band for the whole SNR; the two dashed (blue) areas show the extent of the Fermi 68\% confidence band for each half-disk of the SNR.
\label{fig:fermi_spectrum} }
\end{figure}

The $Fermi$-LAT spectral points for Puppis~A were obtained by dividing the 200 MeV -- 100 GeV 
range into seven logarithmically-spaced energy bins and performing a maximum likelihood spectral 
analysis in each interval, assuming a power-law spectral shape with fixed photon index $\Gamma = 2$ and the X-ray spatial model for Puppis~A. Spectral points of Puppis~A above 3 GeV were determined using the all-phase interval which is allowed by the improved PSF at those energies as well as the lower energy flux of the Vela pulsar. These spectral energy points are in agreement with those derived in the off-pulse window. The result and the butterfly from the overall spectral fit, renormalized to the total phase interval, are presented in Figure~\ref{fig:fermi_spectrum} along with the butterflies corresponding to the fit of the two half-disks of the SNR. The errors on the spectral points represent the statistical and systematic uncertainties added in quadrature. Two main systematic errors have been taken into account: imperfect modeling of the Galactic diffuse emission and uncertainties in the effective area calibration. The first one was estimated by changing artificially the normalization of the Galactic diffuse model by $\pm$6\% as done in \cite{fermi:vela2}. The second one is estimated by using modified IRFs whose effective areas bracket the nominal ones. These bracketing IRFs are defined by envelopes above and below the nominal energy dependence of the effective area by linearly connecting differences of (10\%, 5\%, 20\%) at $\log_{10}(E/{\rm MeV})$ of (2, 2.75, 4), respectively. 

Using the best-fit spatial and spectral model (D1) renormalized to the all-phase interval, 
the $\gamma$-ray luminosity of Puppis A in the 1--100 GeV band is calculated to be 2.7$\times$10$^{34}$ ($D / 2.2$ kpc)$^2$ erg s$^{-1}$. This is among the faintest reported SNRs detected by $Fermi$-LAT. By comparison, SNRs which are interacting with dense clouds, such as IC 443 and W51C have luminosities of $\ge$10$^{35}$ erg s$^{-1}$ \citep{fermi:ic443,fermi:w51c}. SNR W49B is thought to have an age of $\sim$4 kyrs, comparable to Puppis A, yet has a luminosity of 9.3$\times$10$^{35}$ erg s$^{-1}$ above 1 GeV and is interacting with a dense molecular cloud \citep{fermi:w49b}. The relatively low GeV luminosity of Puppis A may be due to either less gas in the vicinity of the SNR, or to an emission mechanism  different than in high luminosity SNRs. We explore this further in \S~\ref{sec:nonthermal} by fitting simple nonthermal emission models.

\subsection{High-Frequency Radio Spectrum}
\label{radio}

\begin{deluxetable}{crrrrrrrrrrr}
\tablecaption{WMAP Flux Density for Puppis A. Five bands are analyzed with effective central frequencies ($\nu_{\rm eff}$) of 23 to 93 GHz (see \S~\ref{radio}). \label{tbl:wmap}}
\tablewidth{0pt} 
\tablehead{ \colhead{Band} & \colhead{$\nu_{eff}$} & \colhead{Flux Density} & \colhead{FWHM} \\ & \colhead{(GHz)} & \colhead{(Jy)} & \colhead{(\degr)}}
\startdata
K	&22.7	&25.8$\pm$2.3 & 0.93 \\
Ka	&33.0	&20.8$\pm$1.7 & 0.68 \\
Q	&40.6	&17.8$\pm$1.5 & 0.53 \\
V	&60.5	&12.3$\pm$1.5 & 0.35 \\
W	&93.0	&3.0$\pm$2.3\tablenotemark{a} & 0.23 
\enddata
\tablenotetext{a}{The source is not detected at 93 GHz. We find a 2$\sigma$ upper limit of 7.6 Jy.} 
\end{deluxetable}

The radio morphology of Puppis A is that of a bright shell which greatly overlaps but is not entirely correlated with the X-ray and IR morphologies. The integrated radio spectrum between 19 MHz and 8.4 GHz is well fit by a single power-law with slope $\alpha$=--0.52$\pm$0.03 \citep{Castelletti2006}. The high resolution VLA observations at 1425 and 327 MHz permitted a detailed study of the radio morphology and spectral index of Puppis A \citep{Castelletti2006}. While it is difficult to quantify, spectral flattening, to an index $\alpha\sim$--0.4, is observed for the radio-bright regions where interaction with denser gas is thought to take place. In contrast, the periphery of the SNR shell appears to show a steeper spectral index of $\alpha\sim$--0.6. The resulting global spectral index is therefore an average of these regions.

\begin{figure}
\centering
\includegraphics[width=6.5in]{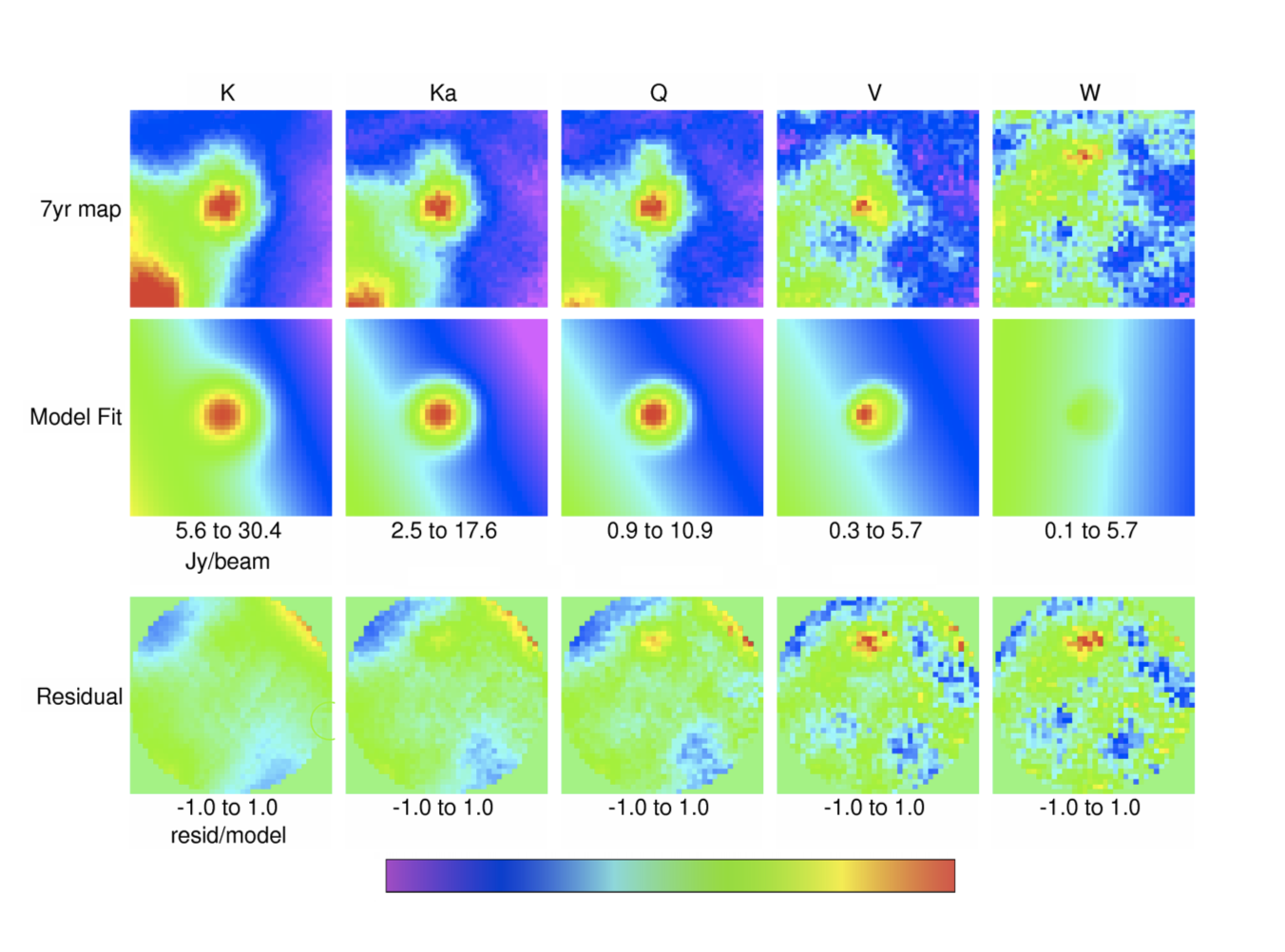}
\caption{Template fits to the WMAP data within a 2$\degr$ radius of Puppis A. Five columns are labeled with the corresponding WMAP radio bands. Three rows present the 7 year skymap images, models resulting from a fit of the 1.4 GHz radio template plus a sloping planar baseline, and the relative residuals (defined as the fit residual divided by the model). The upper and lower limits of the linear colorbar are given beneath each image. Data and models are on the same scale. Puppis A is clearly detected in four bands, but not in the highest frequency W-band (93 GHz). We note that the radio point source to the north of Puppis A, which is clearly visible in the residuals, corresponds to the massive star forming region, IRAS 08212-4159. Including this source in our template fit does not affect the fitted values for the SNR. All images are in celestial coordinates.
\label{fig:wmap} }
\end{figure}

We have used the 7-year all-sky data of the {\it Wilkinson Microwave Anisotropy Probe} (WMAP) to extend the radio spectrum of Puppis A to higher frequencies. Five bands are analyzed with effective central frequencies ($\nu_{\rm eff}$) of 23 to 93 GHz \citep{Jarosik2011}.  To obtain WMAP flux densities we used template fitting of the flux-corrected 1425 MHz radio image \citep{Castelletti2006}. This template image was smoothed with the WMAP beam profiles, and then fit in each band with a constant times the smoothed template plus a sloping planar baseline. To avoid emission from nearby Vela-X PWN, the template fit is restricted to a circular region within a 2\degr\ radius. The 7-year skymaps, fit template and residuals are presented in Figure \ref{fig:wmap}. Fluxes are given in Table \ref{tbl:wmap} with errors estimated from the RMS of the fit residuals in each band. For each band we list the effective Gaussian beam full width at half maximum (FWHM). At 23 GHz (K-band) the beam width is comparable to the angular diameter of the SNR. However, at higher frequencies the remnant is resolved as a spatially extended source. There are no previous radio detections of Puppis A at such high frequencies.

\begin{figure}
\centering
\includegraphics[width=5in]{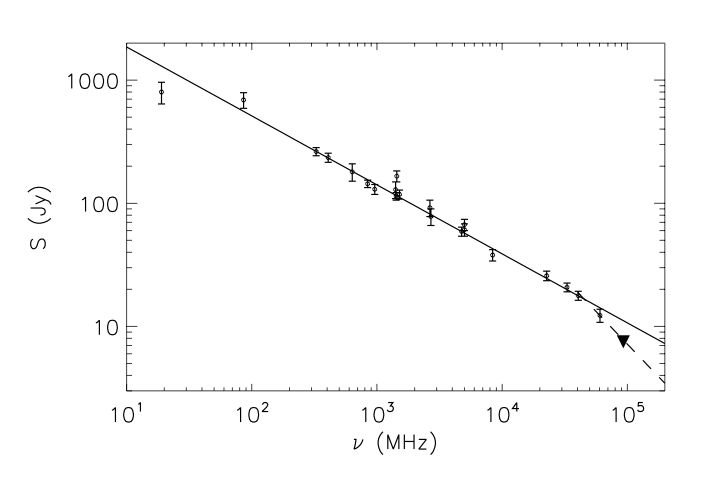}
\caption{Radio spectrum of Puppis A from the flux densities listed in Table \ref{tbl:wmap} and values compiled by \citet{Castelletti2006}. The solid line shows the spectrum for the entire SNR with a flux density of 130 Jy at 1 GHz and a spectral index of $\alpha$=--0.56. The filled triangle shows the 2$\sigma$ upper limit at 93 GHz. The inclusion of WMAP data shows a clear spectral break for frequencies above $\sim$40 GHz. The dashed line shows a spectral break at 40 GHz from an index of --0.55 to --1.05 which is consistent with the WMAP data. 
\label{fig:radio_spectrum} }
\end{figure}

The full radio spectrum of Puppis A from 19 MHz to 93 GHz is shown in Figure \ref{fig:radio_spectrum}. To fit the radio spectrum, we exclude data below 300 MHz, which may suffer from low-frequency absorption by thermal electrons along the line of sight. We find a best-fit 1 GHz flux density of 141$\pm$4 Jy and a radio spectral index of --0.56$\pm$0.01. A radio index of --0.55 is equivalent to a nonthermal particle distribution index of 2.1, matching the spectral index obtained with the $Fermi$-LAT data in \S~\ref{spec}.

{ 
The WMAP data above 40 GHz appear to show a decreasing flux, indicative of a spectral break or cutoff. Both the measured flux density at 63 GHz and the 2$\sigma$ upper limit at 93 GHz fall below the best-fit radio spectrum. WMAP has excellent inter-band calibration \citep{Weiland2011}, so this offers intriguing evidence for spectral curvature at high radio frequencies. To assess the statistical significance of the putative spectral break, we assume a synchrotron cooling break by $\Delta$$\alpha$=--0.5, and apply the F-test to compare the $\chi^{2}$ fit to that of a simple power-law. We find a 2.8$\sigma$ significance for a break at 40 GHz, which is not enough to claim a detection, but highly suggestive. Figure \ref{fig:radio_spectrum} shows that the fit with a spectral break at $\nu_{b} \sim$40 GHz can reproduce the observed data. This is in contrast to the LAT spectrum, where there is no evidence for a break or cutoff at high energies. We discuss the implications of a high-frequency radio break in \S~\ref{discussion}. 
}

\section{Nonthermal Modeling  \label{sec:nonthermal}}

Determining the mechanism responsible for $\gamma$-ray emission is crucial in order to measure the underlying relativistic particle population accelerated by the SNR. 
The excellent correlation between the GeV and X-ray morphologies is suggestive of potential $\gamma$-ray emission mechanisms. First, the shock heated thermal X-ray gas produces strong IR emission by heating dust grains \citep{Arendt2010}. 
Relativistic electrons that inverse Compton (IC) scatter off this IR emission, as well as off the cosmic microwave background can contribute to $\gamma$-ray emission. 
Alternatively, the increasing X-ray brightness from the South-West to the North-East of the remnant is explained by an increasing density of swept-up gas, rising from 0.5 to 4 cm$^{-3}$. The correlation with gas density suggests either nonthermal Bremsstrahlung emission by relativistic electrons or neutral pion decay by relativistic protons, as both scale with increasing density. 

Here we adopt the simplest possible assumption that all emission originates from a single population of accelerated protons and electrons contained in a region characterized by a constant matter density and magnetic field strength. The emitting region is assumed to be homogeneous with a volume given by $V$ = $f$(4$\pi$/3) $R^3$ where $f\le$ 1 denotes a filling factor and $R$ = 15 pc is the radius of the remnant. For simplicity, the filling factor is fixed at $f = 1$ in the following. The particle spectra are assumed to follow a power-law with an exponential cutoff dN/dE $\propto$ $\eta_{e, p}$ E$^{-\Gamma}$ $\times$ exp(-E/E$_{\rm max}$), with the same spectral index and energy cutoff for both electrons and protons. Electrons suffer energy losses due to ionization, Bremsstrahlung, synchrotron processes and IC scattering. The modification of the electron spectral distribution due to such losses was calculated according to \cite{Atoyan1995}, where electrons are assumed to be injected at t = 0 from an impulsive source. The particle spectrum is then evolved for 3700 years, the age of Puppis A.

\begin{figure}
\centering
\includegraphics[height=2.3in]{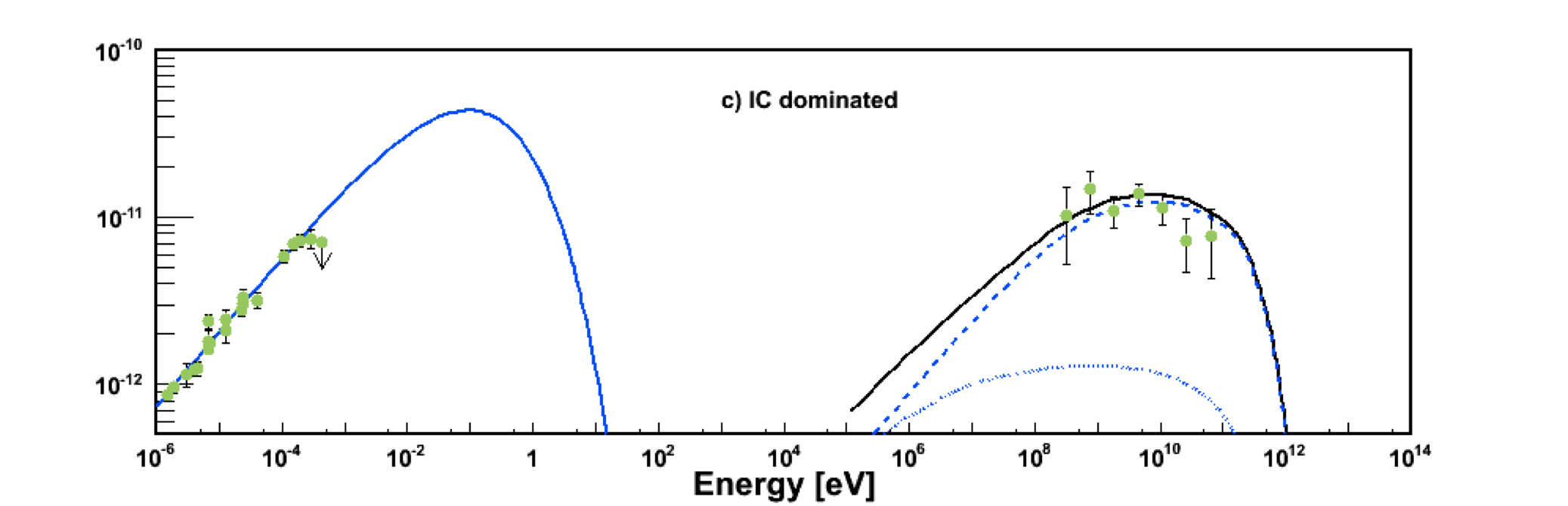}
\includegraphics[height=2.3in]{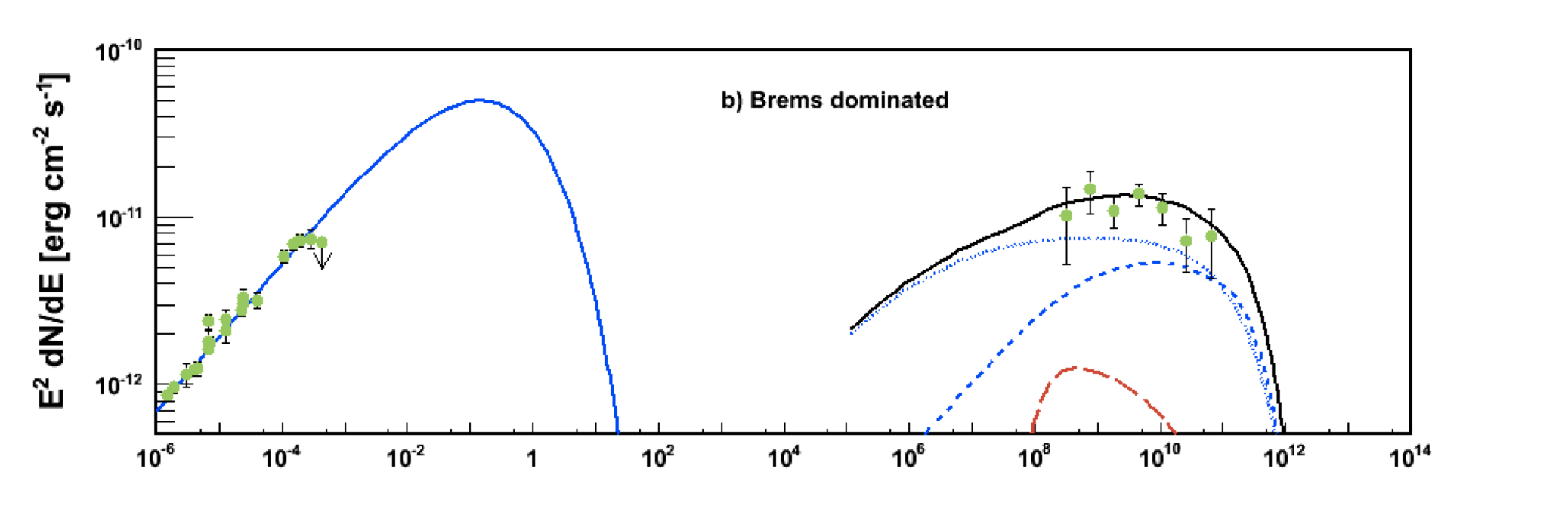}
\includegraphics[height=2.3in]{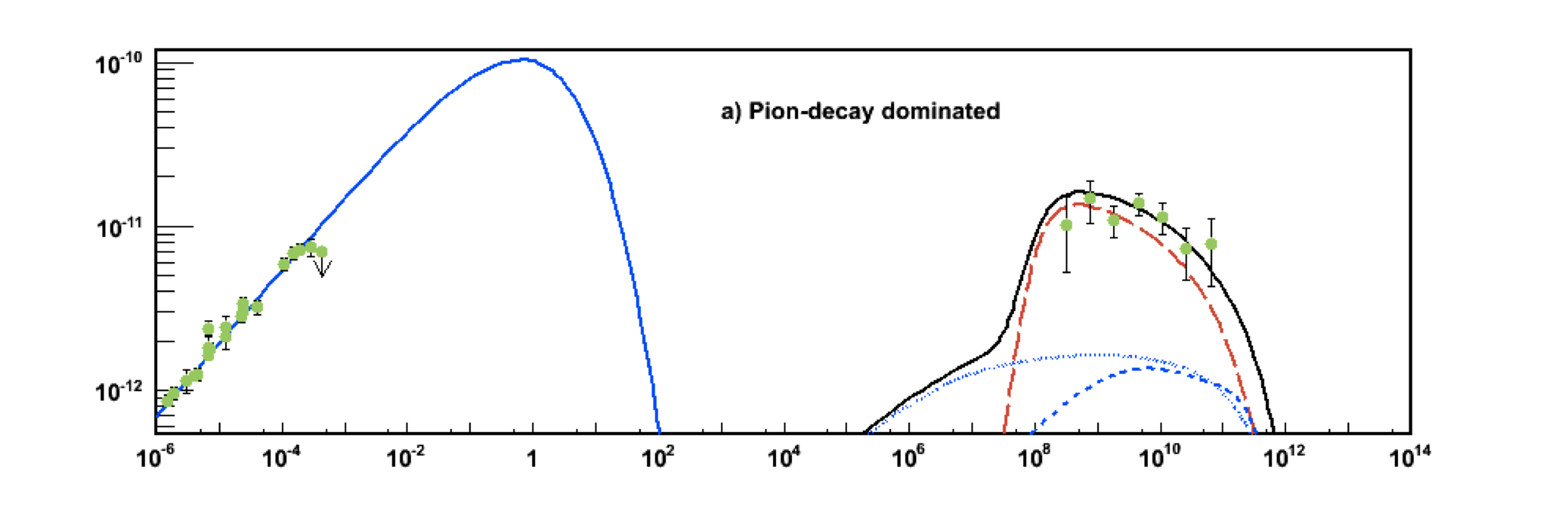}
\caption{SED models of both the $Fermi$-LAT spectrum in Figure \ref{fig:fermi_spectrum} and the radio spectrum in Figure \ref{fig:radio_spectrum}, for which IC (top), Bremsstrahlung (middle) and $\pi^0$-decay  (bottom) are the dominant emission mechanism (see Table~\ref{tbl:sed_models}). In each model the radio data are fit with a synchrotron component. All models show the contributions of $\pi^0$-decay (long dashed, red), Bremsstrahlung (dotted, blue), and IC emission (dashed, blue) from CMB, IR dust photon field, and stellar optical photons. The sum of the three $\gamma$-ray components is shown as a solid black curve. { All data points are shown with 1$\sigma$ error bars. The downward arrow shows the 2$\sigma$ upper limit from WMAP at 93 GHz.}
\label{fig:sed_model}
}
\end{figure}

We consider three models in which each of the three plausible emission mechanisms (IC scattering, Bremsstrahlung, proton-proton interaction) dominates. In fitting the LAT spectrum of Puppis A, the normalization, maximum energy cutoff and magnetic field are left as free parameters. Since we assume that protons and electrons have identical injection spectra, the spectral index of the particle spectrum below the cutoff energy is determined and fixed at $\sim2.1$ by modeling the radio spectrum as synchrotron radiation by relativistic electrons. The spectral break in the WMAP data above 40 GHz being marginal with the current statistics, we do not try to reproduce this break in the following scenarios proposed.  Modeled spectral energy distributions (SEDs) are presented in Figure \ref{fig:sed_model}. Model parameters are given in Table \ref{tbl:sed_models}. We also give the total energy of accelerated particles integrated above 1 GeV for protons, and above 511 keV for electrons. While the chosen parameters are not unique in their ability to fit the broadband spectrum, they are representative of the difficulties and advantages of each scenario. We discuss the viability of each emission model in the following subsections.

\begin{deluxetable}{lcccccccccc}
\tablecaption{One-Zone Model Parameters \label{tbl:sed_models}}
\tablewidth{0pt}
\tablehead{ 
\colhead{Model} &
\colhead{Index} &
\colhead{$E_{\rm max}$} &
\colhead{$n_{\rm H}$} &
\colhead{$B_{\rm tot}$} &
\colhead{$\eta_e$/$\eta_p$} &
\colhead{$W_{p}$} &
\colhead{$W_{e}$}
\\
&
&
\colhead{[TeV]} &
\colhead{[cm$^{-3}$]} &
\colhead{[$\mu$G]} &
&
\colhead{[erg]} &
\colhead{[erg]}
}
\startdata
IC			&2.1 & 0.5 &0.3 	& 8   	&1	 & 8.0$\times$10$^{48}$ 	& 2.9$\times$10$^{49}$ 	\\ 
Brems.		&2.1 & 0.5 &4	& 13		&1 	 & 3.5$\times$10$^{48}$	& 1.3$\times$10$^{49}$	\\ 
$\pi^0$-decay	&2.1	& 0.8 	&4	& 35     	&0.02 &4.0$\times$10$^{49}$	& 2.8$\times$10$^{48}$	\\
\enddata
\end{deluxetable}

\subsection{Inverse Compton Dominated Model}

Puppis A has several photon fields which can be up-scattered by relativistic electrons to produce IC emission: the cosmic microwave background radiation (CMBR), the infrared continuum produced by the collisional excitation of swept-up dust, and the local interstellar radiation field (ISRF). The luminosity of IC $\gamma$-rays is proportional to n$_{ph}$ $\epsilon^{1/2}$, where n$_{ph}$ is the average photon density and $\epsilon$ is the average photon energy \citep{Gaisser1998}. Typically, the CMBR is the dominant radiation field, characterized by n$_{ph} \approx$ 400 cm$^{-3}$ and $\epsilon \approx$ 6.3$\times$10$^{-4}$ eV. Dust emission from Puppis A is fit by a two temperature black body with T$_1$=150 K, n$_{ph}$=12 cm$^{-3}$, $\epsilon$=0.04 eV; and T$_2$=45 K, n$_{ph}$=20 cm$^{-3}$, $\epsilon$=0.01 eV \citep{Arendt1991}. Summing these components, the total contribution to IC $\gamma$-rays from IR dust emission is only $\sim$60\% that from the CMBR. As Puppis A lies at a Galactocentric radius comparable to the Sun, we assume the local ISRF spectrum \citep{Mathis1983} which peaks at 1.2 eV, and contributes $\sim$25\% compared to the CMBR. While shocked dust and the stellar optical background contribute significantly to IC emission, they do not dominate over the CMBR, and cannot solely explain emission from Puppis A.
 
Accounting for these photon fields, an IC-dominated model with a particle index of 2.1 and a cutoff at 500 GeV provides a good fit to the data. With a magnetic field of $\sim$8 $\mu$G, this scenario requires a total energy in CR electrons of 2.9$\times$10$^{49}$ erg. It should be noted that the electron-to-proton ratio is not well constrained and was fixed here at $\eta_{e}$/$\eta_{p} =1$. However, this value needs to be larger than 0.1, and thus in excess of the ratio found for local cosmic-ray abundances, to inject a reasonable energy content in radiating electrons. In the same way, the average gas density needs to be lower than 0.3 cm$^{-3}$ to reduce the Bremsstrahlung component. This limit is very close to the lower limit from thermal X-ray observations \citep{Hwang2005} and is somewhat low for unshocked ISM in the vicinity of a molecular cloud.

\subsection{Bremsstrahlung Dominated Model}

Bremsstrahlung emission, from CR electrons encountering gas of moderate density, can provide a reasonable emission mechanism. X-ray and IR observations constrain the density in the shock interaction region to be $\sim$4 cm$^{-3}$ \citep{Arendt2010}. 
Assuming an electron-to-proton ratio of 1 and a gas density of 4 cm$^{-3}$, the best-fit Bremsstrahlung dominated model has a particle index of 2.1 and a cutoff above 500 GeV. The total energy in CR electrons is 1.3$\times$10$^{49}$ erg, and the magnetic field is $\sim$13 $\mu$G. Given that the SNR is adiabatically expanding, compressing  the swept-up gas and magnetic field lines by a factor of $\sim$4, a magnetic field of 13 $\mu$G is reasonable for compression alone in the post-shock region from which radio and $\gamma$-ray emission may arise. We also note that at densities of $\sim$4 cm$^{-3}$, Bremsstrahlung emission dominates over hadronic $\pi^0$ decay for electron-to-proton ratios greater than $\sim$0.1, again in excess of cosmic-ray abundances.

\subsection{Hadron Dominated Model}

We then consider a $\pi^0$-decay model to account for the broadband spectrum of Puppis~A. The flux and spectrum of the $\gamma$-rays produced by $\pi^0$-decay are calculated following the analytical approximations by \cite{Kelner2006}. Here, we assume a gas density of 4 cm$^{-3}$, and an electron-to-proton ratio of 0.02, consistent with the locally observed cosmic-ray composition. The particle index is fixed at 2.1 with a cutoff above 800 GeV. The total energy in CR hadrons is 4.0$\times$10$^{49}$ erg, and for electrons is 2.8$\times$10$^{48}$ erg. Assuming a typical explosion energy for Puppis A, this hadronic model indicates that a few percent of the initial kinetic energy is transferred to CRs within a few thousand years. 
The magnetic field value required to reproduce the synchrotron radio observations is $\sim$35 $\mu$G, assuming that the radio and $\gamma$-ray emissions arise from the same volume. However, if the $\gamma$-ray emission originates from higher density gas with a lower volume filling fraction than the radio emission, the average magnetic field would be less than our fitted value.

From the three emission models above, it is clear that leptonic scenarios require an electron-to-proton ratio at least ten times larger than the standard cosmic-ray abundance ratio. In addition, the inverse-Compton dominated model requires an extremely low density as well as a relatively low magnetic field value. However, with the current statistics, it is not possible to discriminate between these different models, and the parameters listed in Table \ref{tbl:sed_models} provide an estimate of the energetic particle population in the SNR Puppis A under reasonable assumptions. It should be noted that the CR content (electrons and protons) estimated in the leptonic models should be taken as a lower limit since the electron-to-proton ratio is fixed at a maximum value of $\eta_{e}$/$\eta_{p} =1$ in the fit.

\section{Discussion} \label{discussion}

Puppis A is adiabatically expanding into the interstellar medium in the vicinity of a large molecular cloud. At an age of $\sim$3700 years, the SNR has not yet encountered the bulk of the large cloud, but is likely to at some point in the future. 
This makes Puppis A an interesting transitional case between young SNRs still evolving into a circumstellar medium (e.g. Cas~A), and older SNRs which are interacting with large, dense molecular clouds (e.g. IC 443). At an age of only 350 years, Cas~A is detected at both GeV and TeV energies with a $\gamma$-ray spectral index of $\sim$2.1 \citep{fermi:casa,Acciari2010}. Multi-zone modeling of the SED found that a hadronic CR content of at least 6$\times$10$^{49}$ erg is required \citep{araya10}. This is comparable to the total energy in nonthermal protons in Puppis A estimated in our hadronic model. In contrast, SNR IC 443 has a luminosity one order of magnitude greater than Puppis A, with a total energy content in protons of only $\sim$few times 10$^{49}$ erg (in hadronic models), similar to Puppis A. The difference in luminosity can be easily explained by the much higher average density in IC 443, as it is interacting with large molecular clouds. Cas~A is thought to have a target density comparable to that for Puppis A, and so a comparable luminosity is to be expected.

It is interesting to speculate on how future observations may constrain the nature of the $\gamma$-ray emission in Puppis A: either with improved statistics from $Fermi$-LAT or from a detection at TeV energies. { Extrapolating from the LAT flux and index, the uniform disk has an estimated photon flux of 2$\times$10$^{-11}$ ph cm$^{-2}$ s$^{-1}$ above 0.1 TeV, equivalent to $\sim$5\% of the Crab flux. The current generation of Cherenkov telescopes should be capable of detecting Puppis A, if no high energy cutoff is present. 
} 
If TeV emission is detected toward Puppis A, we expect it to be brightest in the East where there is also enhanced radio and X-ray emission from interaction with higher density ISM toward the adjacent molecular cloud. As can be seen in Figure 2, the centroid of LAT emission moves toward the East with increasing energy. The differences in $\gamma$-ray photon index agree with variations in the observed radio spectral index \citep{Castelletti2006}. The bright Eastern shell appears flatter ($\alpha\sim$--0.5, $\Gamma\sim$2), while the fainter Western extent is generally steeper ($\alpha\sim$--0.7, $\Gamma\sim$2.4). TeV observations can provide an important test 

Future radio observations are also needed to confirm a spectral break or cutoff at $>$40 GHz indicated by the WMAP spectrum. 
Few SNRs have been convincingly shown to have breaks at high frequencies. One such SNR is S147, which has a break at $\sim$1.5 GHz. It was also recently identified as a spatially extended $Fermi$-LAT source \citep{fermi:s147}. To explain the radio and $\gamma$ ray spectra, a two-zone model was invoked: diffuse, low density gas dominates the radio emission at low frequencies, while high-density shock-compressed filaments give rise to $\gamma$-rays and radio emission at frequencies above the radio break.

However, there are significant differences between S147 and Puppis A. S147 is an order of magnitude older and has entered the radiative phase (indicated by dense H$\alpha$ filaments), while Puppis A is still non-radiative, except for a few bright X-ray knots. It is also more difficult to explain the break frequency in Puppis A. For a magnetic field of B$_{\rm \mu G}$, an electron radiating synchrotron emission at a peak frequency $\nu_{\rm GHz}$ has an energy E = 14.7 ($\nu_{\rm GHz}$/B$_{\rm \mu G}$)$^{0.5}$ GeV \citep{Reynolds2008}. The radio break in S147 occurs at $\sim$1.5 GHz, and may be explained by synchrotron cooling over the long lifetime of the remnant. It could also be the effect of damping of magnetohydrodynamic
turbulence (due to ion-neutral collisions) in a model where shock re-accelerated CR particles are adiabatically compressed and energized as a result of radiative cooling behind the cloud shock \citep{Malkov2011}. No radio break is observed for the denser optical filaments. A synchrotron break at $\sim$40 GHz would require a magnetic field in excess of 1 mG to have existed for the entire lifetime of Puppis A, in order to cool the electron spectrum.  However, no corresponding break is observed in the LAT spectrum with the current statistics, so two electron populations would be required to fit the radio and $\gamma$-ray data.

Clearly, further observations to characterize the radio and $\gamma$-ray spectra are needed. High-frequency radio observations are needed to spatially resolve the radio spectral break or cutoff. Increased sensitivity with the continued observations by $Fermi$-LAT will allow any differences between the Eastern and Western regions to be firmly established.
Observations at TeV energies of Puppis A will be interesting in several respects: (1) to determine whether a spectral cutoff or break is present at higher energies and perhaps discriminate between emission mechanisms, (2) to measure differences between East and West of the SNR, and (3) to search for escaping CRs which may have encountered the nearby Eastern cloud.

\section{Conclusions}

We report on high energy $\gamma$-ray emission from the Puppis A SNR detected with $Fermi$-LAT. The source is clearly spatially extended at energies $>$ 800 MeV. The morphology of GeV emission is well correlated with X-ray and IR morphologies.  Spectrally, the SNR is well described by a power-law index of 2.1 in the LAT energy range. 
We also report the detection of Puppis A with WMAP at 20 to 93 GHz. Extending the radio spectrum to high energies reveals a putative spectral break or cutoff at a frequency above 40 GHz, which would have interesting implications for the nonthermal electron population if confirmed. We fit the radio to $\gamma$-ray SEDs using IC, Bremsstrahlung and hadronic dominated models. All emission mechanisms are able to fit the data, though with different magnetic field strengths and energetics of relativistic particles. For the hadronic models a total energy in CR protons of $\sim$4$\times$10$^{49}$ erg is needed, whereas for leptonic models, at least (1 -- 3)$\times$10$^{49}$ erg in relativistic electrons is required.
While Puppis A has a relatively low $\gamma$-ray luminosity in comparison to other SNRs identified by $Fermi$, this may be explained by the low average gas density, n$_{\rm H}\sim$4 cm$^{-3}$, swept-up by the SNR. Future observations at high radio frequencies and at higher $\gamma$-ray energies will help to differentiate between leptonic and hadronic emission models.

\acknowledgements
The \textit{Fermi} LAT Collaboration acknowledges generous ongoing support
from a number of agencies and institutes that have supported both the
development and the operation of the LAT as well as scientific data analysis.
These include the National Aeronautics and Space Administration and the
Department of Energy in the United States, the Commissariat \`a l'Energie Atomique
and the Centre National de la Recherche Scientifique / Institut National de Physique
Nucl\'eaire et de Physique des Particules in France, the Agenzia Spaziale Italiana
and the Istituto Nazionale di Fisica Nucleare in Italy, the Ministry of Education,
Culture, Sports, Science and Technology (MEXT), High Energy Accelerator Research
Organization (KEK) and Japan Aerospace Exploration Agency (JAXA) in Japan, and
the K.~A.~Wallenberg Foundation, the Swedish Research Council and the
Swedish National Space Board in Sweden.

Additional support for science analysis during the operations phase is gratefully
acknowledged from the Istituto Nazionale di Astrofisica in Italy and the Centre National d'\'Etudes Spatiales in France.

The Parkes radio telescope is part of the Australia Telescope which is funded by the Commonwealth Government for operation as a National Facility managed by CSIRO. We thank our colleagues for their assistance with the radio timing observations.

The WMAP mission is made possible by the support of the Science Mission Directorate Office at NASA Headquarters. This research has made use of NASA's Astrophysics Data System Bibliographic Services, and data obtained from the High Energy Astrophysics Science Archive Research Center (HEASARC), provided by NASA's Goddard Space Flight Center. 

We thank Nils Odegard for assisting us with analysis of WMAP data, and Gloria Dubner for providing us with the 1.4 GHz radio image of Puppis A. JWH acknowledges support by an appointment to the NASA Postdoctoral Program at the Goddard Space Flight Center, administered by Oak Ridge Associated Universities through a contract with NASA, and also by a grant from the $Fermi$ Guest Investigator Program. MHG acknowledges support from the Alexander von Humboldt Foundation. MLG acknowledges funding by contract ERC-StG-259391 from the European Community.

\newpage

\end{document}